\documentclass[prd,amsmath,amssymb,superscriptaddress,nofootinbib,twocolumn]{revtex4}

\usepackage{xcolor}
\usepackage{bm}
\usepackage[normalem]{ulem}
\usepackage{babel}
\usepackage{textcomp}
\usepackage{amstext}
\usepackage{graphicx}
\usepackage{esint}

\usepackage[unicode=true,pdfusetitle,
 bookmarks=true,bookmarksnumbered=false,bookmarksopen=false,
 breaklinks=false,pdfborder={0 0 1},backref=false,colorlinks=true]
 {hyperref}
\hypersetup{
 pdfborderstyle=,linkcolor=blue,citecolor=cyan}

\newcommand{\nn}{\nonumber}

\newcommand{\beq}{\begin{equation}}
\newcommand{\eeq}{\end{equation}}
\newcommand{\bqa}{\begin{eqnarray}}
\newcommand{\eqa}{\end{eqnarray}}

\newcommand{\bseq}{\begin{subequations}}
\newcommand{\eseq}{\end{subequations}}

\makeatletter

\usepackage{babel}
\usepackage{xcolor}
\usepackage{color}

\begin{document}

\title{Azimuthal asymmetry in $J/\psi+\gamma$ and $J/\psi+J/\psi$ production in ultraperipheral heavy-ion collisions at LHC
}

\author{Yu Jia~\footnote{jiay@ihep.ac.cn}}
\affiliation{Institute of High Energy Physics,
Chinese Academy of Sciences, Beijing 100049, China\vspace{0.2cm}}
\author{Wen-Long Sang~\footnote{wlsang@swu.edu.cn}}
\affiliation{School of Physical Science and Technology, Southwest University, Chongqing 400700, China\vspace{0.2cm}}
\author{Xiaonu Xiong~\footnote{xnxiong@csu.edu.cn}}
\affiliation{School of Physics and Electronics, Central South University, Changsha 410083 , China\vspace{0.2cm}}
\author{Jian Zhou~\footnote{jzhou@sdu.ac.cn}}
\affiliation{Key Laboratory of
Particle Physics and Particle Irradiation (MOE),Institute of
Frontier and Interdisciplinary Science, Shandong University,
(QingDao), Shandong 266237, China \vspace{0.2cm}}
\affiliation{Southern Center for Nuclear-Science Theory (SCNT), Institute of Modern Physics, Chinese Academy of Sciences, HuiZhou, Guangdong
516000, China\vspace{0.2cm}}
\author{Ya-jin Zhou~\footnote{zhouyj@sdu.ac.cn}}
\affiliation{Key Laboratory of
Particle Physics and Particle Irradiation (MOE),Institute of
Frontier and Interdisciplinary Science, Shandong University,
(QingDao), Shandong 266237, China \vspace{0.2cm}}

\date{\today}

\begin{abstract}
Two-photon collision in ultraperipheral heavy-ion collisions (UPCs) provides a unique and powerful platform 
for probing QCD with linearly polarized quasi-real photons. While photon polarization effects have been recognized
in dilepton and even in light hadrons production, their consequences for heavy quarkonium production remain unexplored. 
 In this work we investigate for the first time the $\gamma\gamma \to J/\psi+\gamma(J/\psi)$ channels in Pb-Pb UPCs at 
the Large Hadron Collider (LHC), by integrating the non-relativistic QCD (NRQCD) factorization approach with 
the transverse-momentum-dependent (TMD) photon distributions. 
  Based on the helicity amplitudes at lowest order in strong coupling and velocity expansion, 
we predict sizable $\cos(2\phi)$ and $\cos(4\phi)$ azimuthal asymmetries arising from the interference of 
 linearly polarized photon states. 
These azimuthal-dependent observables, defined as the ratios of weighted to unweighted cross sections, are expected to be stable 
against including the higher-order radiative corrections and varying nonperturbative NRQCD matrix elements, 
thus offering a fresh test of quarkonium production mechanism and the photon TMD structure in the ultrarelativistic limit.
\end{abstract}

\maketitle

\section{Introduction}

Ultraperipheral heavy-ion collisions (UPCs) at the Large Hadron Collider ({\tt LHC}) and Relativistic Heavy Ion Collider ({\tt RHIC}) 
generate intense electromagnetic fields that serve a potent source of quasi-real photons. 
These coherently emitted photons are linearly polarized along the direction of their transverse momentum, 
a property that can induce significant azimuthal asymmetries in final-state particle distributions. 
A prominent example is the sizable $\cos(4\phi)$ modulation observed by the {\tt STAR} collaboration 
in the Breit-Wheeler process $\gamma \gamma \to l^+ l^-$~\cite{STAR:2019wlg}, 
with further investigations carried out by {\tt ALICE}~\cite{ALICE:2019tqa,ALICE:2020ugp,ALICE:2021gpt,ALICE:2013wjo}, 
{\tt CMS}~\cite{CMS:2020skx}, {\tt ATLAS}~\cite{ATLAS:2020epq}, and {\tt STAR}~\cite{STAR:2022wfe}. 
These measurements underscore the potential of utilizing linearly polarized photons to probe Quantum Chromodynamics (QCD) and Quantum Electrodynamics (QED) in a clean environment, largely free from hadronic background~\cite{Li:2019sin,Li:2019yzy}. 
This has stimulated broad theoretical interest in two-photon processes, ranging from studies of QED in extreme conditions~\cite{Hagiwara:2020juc,Hagiwara:2021xkf,Xing:2020hwh,Mantysaari:2023prg,Mantysaari:2022sux,Brandenburg:2024ksp,Lin:2024mnj,Shao:2024nor,Hagiwara:2017fye,Iancu:2023lel,Niu:2022cug,Jiang:2024vuq,Klein:2016yzr,Bertulani:1987tz,Bertulani:2005ru,Baltz:2007kq,Zhao:2022dac,Copinger:2018ftr,Klein:2018fmp,Klein:2020jom,Shao:2023zge,Shao:2022stc,Hu:2024bsm,Zhang:2024mql,Hencken:2004td,Zha:2018tlq,Brandenburg:2021lnj,Xiao:2020ddm,Wang:2021kxm,Wang:2022gkd,Lin:2022flv,Shao:2025oeb} and searches for new physics~\cite{Xu:2022qme,Shao:2023bga}~\footnote{
It is worth mentioning that, a recent preliminary study of {\tt STAR} Collaboration has 
reported the observation of light hadron pair production in UPCs, {\it viz.}, 
$\gamma\gamma\to p \bar{p}$. For some theoretical analysis, 
see Refs.~\cite{Zhang:2024mql,Hu:2024bsm}. }. 

Numerous proposals has been advocated to extracting the gluon TMDs through azimuthal 
asymmetries in high energy $pp$ collisions~\cite{Metz:2011wb,Ma:2014oha,Zhang:2014vmh,Ma:2015vpt,Boer:2016fqd,Boer:2017xpy,Lansberg:2016abn,Zhang:2017uiz,Lansberg:2017dzg,Bacchetta:2018ivt,Scarpa:2019fol,Boer:2020bbd,Kishore:2021vsm,Taels:2022tza,liu:2023unl,Caucal:2023fsf,Maxia:2024cjh,Chen:2025gwp}, 
among which the associated heavy quarkonium production with a photon, {\it e.g.}, 
$gg\to J/\psi + \gamma+X$,  stands out as a clean probe, partly due to clean experimental siguature of $J/\psi$ and photon~\cite{denDunnen:2014kjo,Alimov:2024pqt}, partly 
because the colorless final state minimizes final-state interactions~\cite{Hatta:2021jcd,Hatta:2020bgy}.  

Somewhat surprisingly, despite the great experimental and theoretical cleanness, 
the azimuthal asymmetry in simpler cousin process $\gamma\gamma\to J/\psi + \gamma$ in UPCs, 
and the impact of the photon linear polarization, has never been explored in literature.
To date existing studies of exclusive heavy quarkonium production channels in UPCs, exemplified by
$\gamma\gamma\to J/\psi + \gamma (J/\psi)$~\cite{Qiao:2001wv,Baranov:2012vu,Yang:2020rvo,Yang:2020xkl,He:2024lrb,Yang:2025vcs,Chen:2025gwp,Jia:2025qul}, 
have predominantly relied on the {\it equivalent photon approximation} (EPA), 
which is essentially equivalent to QED collinear factorization, 
combined with non-relativistic QCD (NRQCD) factorization approach~\cite{Bodwin:1994jh}. 
By inherently averaging over the transverse polarization of the incoming photons, the EPA 
approach is unable to capture the azimuthal modulations driven by the interference of helicity states, 
thus leaving a gap in our understanding of quarkonium production mechanisms in the TMD regime.

The aim of this paper is to bridge this gap by combining the QED TMD factorization with 
the NRQCD approach to account for the photon-induced quarkonium production of 
$J/\psi + \gamma(J/\psi)$ in UPCs. 
 Our input is the  helicity amplitudes for $\gamma\gamma \to J/\psi + \gamma(J/\psi)$ 
at lowest-order (LO) in the charm quark velocity, $v$, and QCD coupling constant $\alpha_s$.  
We present the predictions for $\cos(2\phi)$ and $\cos(4\phi)$ azimuthal asymmetries, 
which are direct consequences of the linear polarization of the incoming quasi-real photons. 
Under realistic {\tt LHC} kinematic conditions, we find that these asymmetries are sizable 
with cross sections large enough to yield a significant number of events per run. 
Furthermore, we anticipate that these predicted polarization observables, defined as ratios of weighted to unweighted cross sections, 
are theoretically stable against uncertainties in NRQCD long-distance matrix elements (LDMEs) and 
including the higher order radiative and relativistic corrections.  
  This work may provide a novel window of testing the exclusive quarkonium production mechanism.

\section{$J/\psi+\gamma (J/\psi)$ production in UPCs}

\begin{figure}[htbp]
\begin{center}
\includegraphics[width=0.22\textwidth]{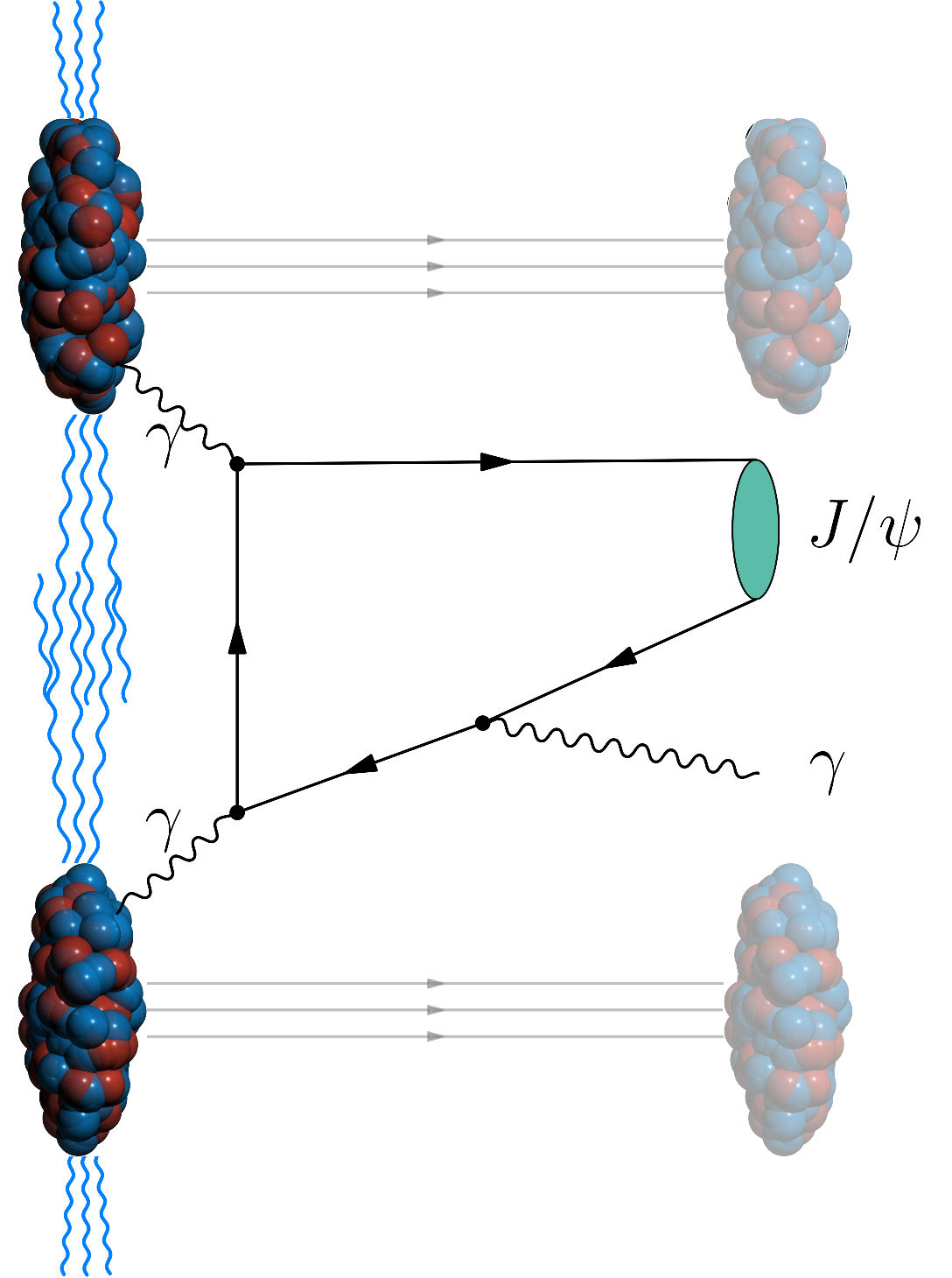} 
\includegraphics[width=0.22\textwidth]{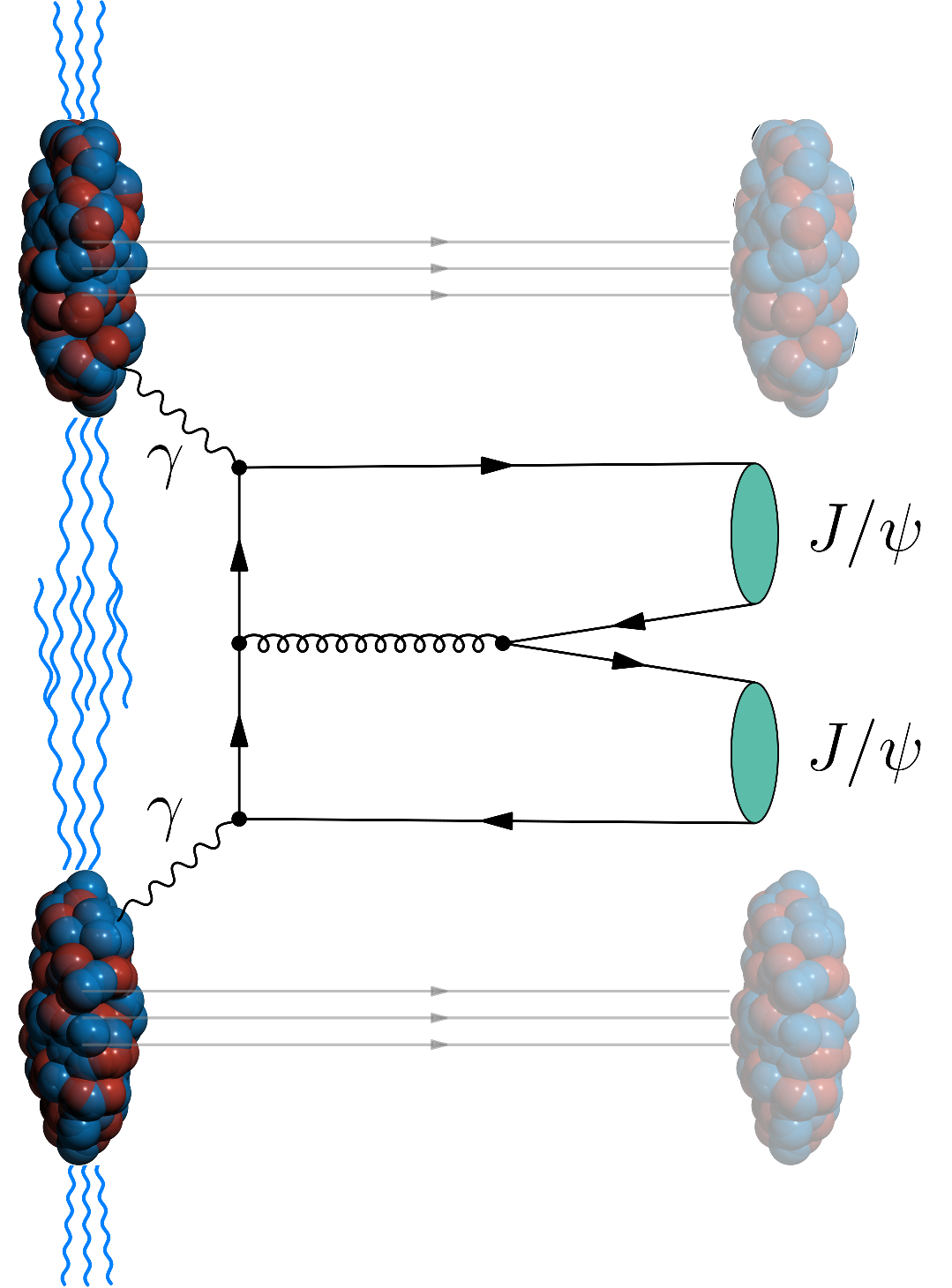} 
\caption{Representative lowest-order diagrams for $\gamma\gamma \to J/\psi + \gamma$ (left) and 
$\gamma\gamma \to J/\psi + J/\psi$ (right). }
\label{fig:feynman}
\end{center}
\end{figure}

The exclusive production of $J/\psi+\gamma(J/\psi)$ in  UPCs is mediated by two-photon collision, 
with both nuclei remain intact. Let us specialize to the reactions 
${\rm Pb} + {\rm Pb}\to {\rm Pb}+{\rm Pb}+J/\psi+\gamma (J/\psi)$.
The kinematics of the partonic subprocesses is specified by
\beq
\gamma(x_{1}P+k_{1\perp})+\gamma(x_{2}\overline{P}+k_{2\perp})  \rightarrow J/\psi(p_{1})+\gamma(J/\psi)(p_{2}),
\eeq
where $x_{1}P$ and $x_{2}\overline{P}$ represent the 
longitudinal momenta of two incoming photons, and $k_{1,2\perp} = (0, 0, \bm{k}_{1,2\perp})$
signify the corresponding transverse momenta of incoming photons.
The representative lowest-order Feynman diagrams for these partonic reactions 
are depicted in Fig.~\ref{fig:feynman}.  

As mentioned before, the quasi-real photons emitted from the relativistic lead ions are linearly polarized,
with the polarization vectors aligned with their transverse momenta $\bm{k}_{1,2\bot}$.
Consequently, to access the azimuthally-dependent observables in the correlation limit, 
one ought to utilize the TMD factorization approach, rather than the standard collinear factorization approach widely used in the preceding studies in two-photon physics. 
To this end, it is convenient to introduce two additional transverse momenta variables: 
${\bm P}_\perp \equiv {{\bm p}_{1\perp}-{\bm p}_{2\perp}\over 2}$, and
${\bm q}_\perp\equiv {\bm p}_{1\perp}+{\bm p}_{2\perp}={\bm k}_{1\perp}+{\bm k}_{2\perp}$.
The azimuthal angle is defined by $\cos\phi \equiv \hat{\bm P}_\perp \cdot \hat{\bm q}_\perp$.
In the correlation limit with $\vert {\bm q}_\perp\vert \ll \vert {\bm P}_\perp\vert$, 
it is legitimate to approximate ${\bm P}_\perp\approx {\bm p}_{1\perp} \approx -{\bm p}_{2\bot}$.

To incorporate both the   impact parameter $\bm{b}_\perp$)   and the $\bm{q}_\perp$ dependence simultaneously,
one can derive the differential cross section for the associated $J/\psi+\gamma (J/\psi)$ production, 
following the formalism originally developed in Refs.~\citep{Vidovic:1992ik,Hencken:1994my}, 
which is valid in the correlation limit. 
After some manipulation, we cast the azimuthal-dependent differential cross section in the following convolutional form \cite{Jia:2024hen}:
\begin{widetext}
\begin{eqnarray}
\frac{d\sigma}{d^{2}\bm{p}_{1\perp}d^{2}\bm{p}_{2\perp}dy_{1}dy_{2}d^{2}\bm{b}_{\perp}} 
&&= \frac{1}{32\pi^{2}Q^{4}}\int d^{2}{\bm k}_{1\perp}\, d^{2}{\bm k}_{2\perp}\, \frac{d^{2}{\bm k}_{1\perp}'}{(2\pi)^{2}}
\delta^{(2)}({\bm q}_{\perp}-{\bm k}_{1\perp}-{\bm k}_{2\perp})\, e^{i({\bm k}_{1\perp}-{\bm k}_{1\perp}')\cdot {\bm b}_{\perp}} 
\nonumber \\
&& \times \Big\{ 
\cos(\phi_{1}-\phi_{2})\,\cos(\phi_{1}'-\phi_{2}')\,|M_{++}|^{2}
+ \cos(\phi_{1}+\phi_{2})\,\cos(\phi_{1}'+\phi_{2}')\,|M_{+-}|^{2} 
\nonumber \\
&& \hspace{0.45cm}
- \cos(\phi_{1}+\phi_{2})\,\cos(\phi_{1}'-\phi_{2}')\,M_{++}M_{+-}^{*} 
- \cos(\phi_{1}-\phi_{2})\,\cos(\phi_{1}'+\phi_{2}')\,M_{+-}M_{++}^{*} 
\Big\}\label{eq:cross section} 
\nonumber \\
&& \times  {\cal F}(x_{1},{\bm k}_{1\perp}^{2})\,{\cal F}^{*}(x_{1},{\bm k}_{1\perp}'^{2})\,{\cal F}(x_{2},{\bm k}_{2\perp}^{2})\,{\cal F}^{*}(x_{2},{\bm k}_{2\perp}'^{2}),
\label{semi-inclusive:production:cross:section}
\end{eqnarray}
\end{widetext}
where  $y_{1,2}$ denote the rapidities of the outgoing photon and $J/\psi$, 
connected with the incident photons' longitudinal momentum fractions $x_{1,2}$ via
$x_{1,2}=\frac{1}{\sqrt{s}}(m_\perp e^{\pm y_1}+P_\perp e^{\pm y_2})
$ for $J/\psi+\gamma$ and
$x_{1,2}=\frac{m_\perp}{\sqrt{s}}( e^{\pm y_1}+ e^{\pm y_2})$ for $J/\psi+J/\psi$,
where $m_\perp=\sqrt{P_\perp^2+m_{J/\psi}^2}$ denotes the transverse mass of the $J/\psi$.
($\sqrt{s}$  denotes the center-of-mass energy of each nucleon pair from the colliding nuclei). 
$\phi_{1,2}$ signify the azimuthal angles between ${\bm k}_{1,2\bot}$ and ${\bm P}_{\bot}$.
${\bm k}'_{1,2\perp}$ denote the transverse momenta of the incoming photons in the conjugated production amplitude,
with $\phi_{1,2}'$ denoting the azimuthal angles between ${\bm k}_{1,2\bot}'$ and 
${\bm P}_{\bot}$~\footnote{The appearance of ${\bm k}'_{i\perp}$ arises from incorporating the impact parameter dependence into our calculation. The variable $\bm{b}_{\perp}$ enters the cross section through the phase factor in Eq.~\eqref{eq:cross section}. Upon integrating over $\bm{b}_{\perp}$ in Eq.~\eqref{eq:cross section}, the cross section, which depends jointly on $\bm{b}_\perp$ and $\bm{q}_\perp$, reduces to the standard result obtained in TMD factorization. For example, see Ref.~\cite{Jia:2024xzx} for a related discussion in a different context, where the lead ion is replaced by a pointlike electron or positron.}.

The nonperturbative distribution function ${\cal F}(x,\bm{k}_{i\perp}^2)$ in Eq.~\eqref{eq:cross section}
characterizes the probability amplitude for finding a photon that carries the prescribed light-momentum fraction and
transverse momenta inside a lead nucleus.
It is intimately related to the standard photon TMD parton distribution functions (PDFs) of a heavy nucleus:
\begin{eqnarray}
\label{photon:TMD:distributions:def}
&& \int \frac{dy^-\, d^2 y_\perp}{P^+ (2\pi)^3} e^{i k \cdot y}
\left\langle A \left| F_{+\perp}^{\mu}(0)\, F_{+\perp}^{\nu}(y) \right| A \right\rangle \Big|_{y^+ = 0} \nonumber \\
&& = \frac{\delta_\perp^{\mu \nu}}{2}\, x f_1(x, {\bm k}_\perp^2)
+ \left( \frac{k_\perp^\mu k_\perp^\nu}{{\bm k}_\perp^2} - \frac{\delta_\perp^{\mu \nu}}{2} \right) x h_1^{\perp}(x, {\bm k}_\perp^2) ,
\end{eqnarray}
where $f_1$ and $h_1^{\perp }$ signify the
unpolarized and linearly-polarized photon TMD distributions, respectively.
The transverse metric tensor in \eqref{photon:TMD:distributions:def} is defined by
$\delta_\perp^{\mu\nu}=-g^{\mu\nu}+ {P^\mu n^\nu+P^\nu n^\mu\over P\cdot n}$ with
$n^\mu=(1,-1,0,0)/ \sqrt{2}$, and $k_\perp^2=\delta_\perp^{\mu\nu} k_{\perp\mu} k_{\perp\nu}$.
At small $x$, the TMD PDFs $f_1$ and $h_{1}^{\bot}$ are simply identified with the square of 
${\cal F}$~\cite{Li:2019sin,Li:2019yzy}:
\beq
\label{eq:photonTMD}
 x f_1(x,\bm{k}_{\perp}^2) = x h_1^\perp(x,\bm{k}_{\perp}^2) =  \left| {\cal F} (x, \bm{k}^2_\perp) \right|^2.
\eeq
 In our actual numerical  calculation,  
${\cal F} (x, \bm{k}^2_\perp)$ is determined by the Woods-Saxon distribution which is taken from \cite{Klein:2016yzr}.

For later use, let us specify the polarization vectors of the photon with definite helicities:
\beq
\epsilon^\mu(k_1,\pm) = \frac{1}{\sqrt{2}}(0,\mp 1,-i,0).
\label{pvector:1:plus:minus}
\eeq
One can define the polarization vector of the second photon to be
$\epsilon^\mu(k_2,\pm)= \epsilon^\mu(k_1,\mp)$, following Jacob-Wick's second particle
phase convention~\cite{Haber:1994pe}.

When evaluating the helicity amplitudes, it is constructive to reexpress the rank-2 tensors in \eqref{photon:TMD:distributions:def}
in terms of the photon's polarization vectors \cite{Jia:2024xzx}:
\begin{subequations}
\bqa
& & \!\!\! \delta_\perp^{\mu \nu} = \epsilon^\mu(k_i,+) \epsilon^\nu(k_i,-)+\epsilon^\nu(k_i,+) \epsilon^\mu(k_i,-),
\\
&& \!\!\! \delta_\perp^{\mu\nu}- 2\frac{k_{i\perp}^\mu
k_{i\perp}^\nu}{k_{i\perp}^2}
 \\&&  = e^{2i\phi_i} \epsilon^\mu(k_i,+) \epsilon^\nu(k_i,+) + e^{-2i\phi_i} \epsilon^\mu(k_i,-) \epsilon^\nu(k_i,-),
\nn
\eqa
\end{subequations}
with $i=1,2$. For definiteness, we have chosen ${\bm P}_\perp$ to align with the $\hat{x}$-axis, 
and
$\phi_1$ and $\phi_2$ represent the azimuthal angles between ${\bm k}_{1\perp}$, 
${\bm k}_{2\perp}$ and ${\bm P}_\perp$.

The coefficient $M_{\lambda_{1},\lambda_{2}}$ in Eq.~\eqref{eq:cross section} 
denotes the helicity amplitude for the partonic reactions 
$\gamma(x_1 P,\lambda_1)\gamma(x_2\overline{P},\lambda_{2})\to J/\psi(p_{1})+\gamma(J/\psi)(p_{2})$.
$M_{\lambda_{1},\lambda_{2}}M_{\lambda_{1}^{\prime},\lambda_{2}^{\prime}}^{*}$
in Eq.~\eqref{eq:cross section} signifies a shorthand for the interference between one helicity amplitude with 
another conjugated one, which bear different photons' helicity configurations, 
yet with all the polarizations of the final-state particles ($J/\psi$ polarizations $\lambda_{3,4}=0,\pm 1$ and photon helicity $\lambda_4=\pm 1$) summed over:
\beq
M_{\lambda_{1},\lambda_{2}}M_{\lambda_{1}^{\prime},\lambda_{2}^{\prime}}^{*} \equiv
 \sum_{\lambda_3} \sum_{\lambda_4} \, M_{\lambda_{1},\lambda_{2},\lambda_3,\lambda_4} M^*_{\lambda_{1}^{\prime},\lambda_{2}^{\prime},\lambda_3,\lambda_4}.
\eeq
Note that the transverse momenta of the incoming photons have been set to zero when calculating the helicity amplitudes, 
in line with the spirit of TMD factorization. Parity invariance has been invoked to reduce the number of encountered  
helicity configurations of incoming photons in Eq.~\eqref{eq:cross section}.

We evaluate all the encountered helicity amplitudes within the NRQCD factorization approach~\cite{Bodwin:1994jh}. 
 At LO in the strong coupling $\alpha_s$ and velocity expansion, the calculation can be facilitated 
by using the covariant projector approach by replacing $J/\psi$ with a comoving color-singlet $c\bar{c}$ pair
carrying the quantum number $^3S_1$. We devote Appendix~A to the explicit expressions of 
all the encountered $M_{\lambda_{1},\lambda_{2}}M_{\lambda_{1}^{\prime},\lambda_{2}^{\prime}}^{*}$
in this work.

Equation~\eqref{semi-inclusive:production:cross:section} constitutes the master formula of this work. 
The first two terms in the curly bracket contribute to the azimuthally averaged cross section,
which differs from the expression for the azimuthally averaged cross section derived in Refs.~\cite{Klusek-Gawenda:2016euz,Harland-Lang:2020veo,Shao:2022cly,AH:2023kor}
due to the intriguing entanglement between the impact parameter and polarization vectors of the coherent photons. 
 More interestingly,  the last two terms in the curly bracket in Eq.~\eqref{eq:cross section}, {\it viz.}, 
the terms proportional to $M_{++} M_{+-}^*$ and $M_{+-} M_{++}^*$,
entail the interference between different helicity amplitudes, which is a direct consequence of the linear polarization of
the incoming quasi-real photons.
After integrating over the transverse momenta of incoming photons, the
angular correlations between ${\bm k}_{i\bot}$, ${\bm k}_{i\bot}'$ and ${\bm P}_{\bot}$ in the last two terms in the curly bracket
are converted into the angular correlation between ${\bm q}_\perp$ and ${\bm P}_\perp$.
Specifically, the interference between the unpolarized and linearly polarized photon distributions generates the $\cos(2\phi)$ asymmetry, while the interference between two linearly polarized photon distributions drives the $\cos(4\phi)$ modulation.

\section{Phenomenology}
To facilitate the comparison between experiment and theory,
we define the average value of $\cos(2\phi)$ and $\cos(4\phi)$ as
\beq
\langle \cos(n\phi) \rangle \equiv \frac{ \int d\sigma \, \cos(n\phi) }{ \int d\sigma },
\label{azimuthal:variable}
\eeq
with $n=2,4$. 

In the numerical calculations, we adopt the following input parameters: 
the center-of-mass energy $\sqrt{s_{\rm NN}} = 5.02$~TeV, the charm quark mass $m_c = 1.5$~GeV, and 
the Schr\"odinger radial wave function at the origin for the $J/\psi$, $|R_{J/\psi}(0)|^2=0.9215$~${\rm GeV}^3$~\cite{Bodwin:2007ga}.
We use the package {\tt HOPPET}~\cite{Karlberg:2025hxk} to evaluate the running strong coupling constant 
$\alpha_s(\mu=Q/2)$ at one-loop accuracy. 
We also impose the following kinematic cuts: $1.6 < |y_1|, |y_2| < 2.4$, $Q < 20$~GeV, and $P_\perp > 200$~MeV. 
In addition, we integrate over $q_\perp$ from 0 to 200~MeV.

\begin{figure}[htbp]
   \begin{center}
   \includegraphics[width=0.45\textwidth]{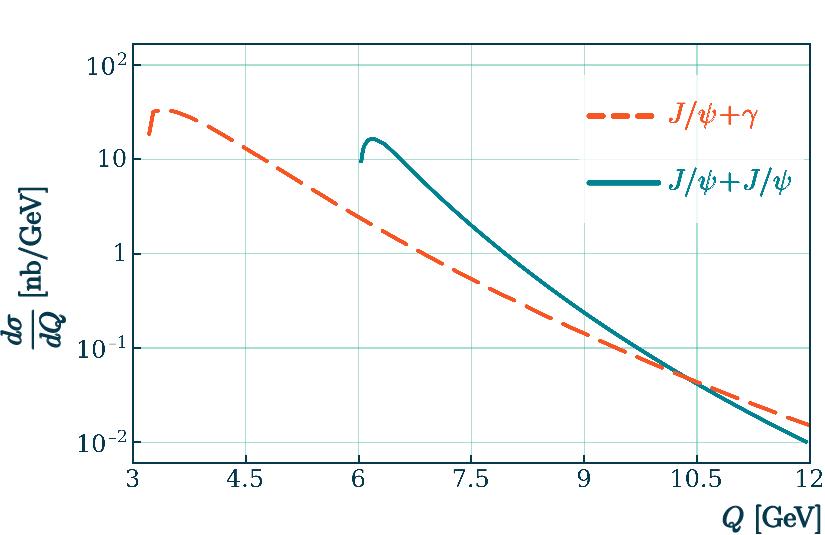}
   \caption{The spectra of the invariant mass of $J/\psi+\gamma(J/\psi)$ for exclusive $J/\psi + \gamma$ and $J/\psi + J/\psi$ 
   production in Pb+Pb UPCs at $\sqrt{s_{\rm NN}} = 5.02$~TeV, under the cuts $1.6 < |y_{1,2}| < 2.4$, $P_\perp > 200$~MeV, with $q_\perp$ integrated up to 200~MeV. }
   \label{fig:ds_dpt}
   \end{center}
   \end{figure}

Our numerical studies indicate that the exclusive production of $J/\psi + \gamma$ and $J/\psi + J/\psi$ in Pb+Pb UPCs 
at $\sqrt{s_{\rm NN}} = 5.02$~TeV yields sizable cross sections within the aforementioned kinematic constraint. 
As shown in Fig.~\ref{fig:ds_dpt}, the differential cross sections $d\sigma/dQ$ as functions of the invariant mass $Q$ 
exhibit similar behaviors for the two processes, both peaking near their respective production thresholds and decreasing with increasing $Q$. 
Remarkably, the $J/\psi + \gamma$ channel features a larger cross section,  despite the nominal suppression brought by the QED fine structure constant. 
Concretely speaking, the integrated cross section for $J/\psi+J/\psi$ production
within the prescribed kinematic window is about $13.4$~nb, 
in contrast with $43.4$~nb  for $J/\psi + \gamma$ production.  

\begin{figure}[htbp]
\begin{center}
\includegraphics[width=0.45\textwidth]{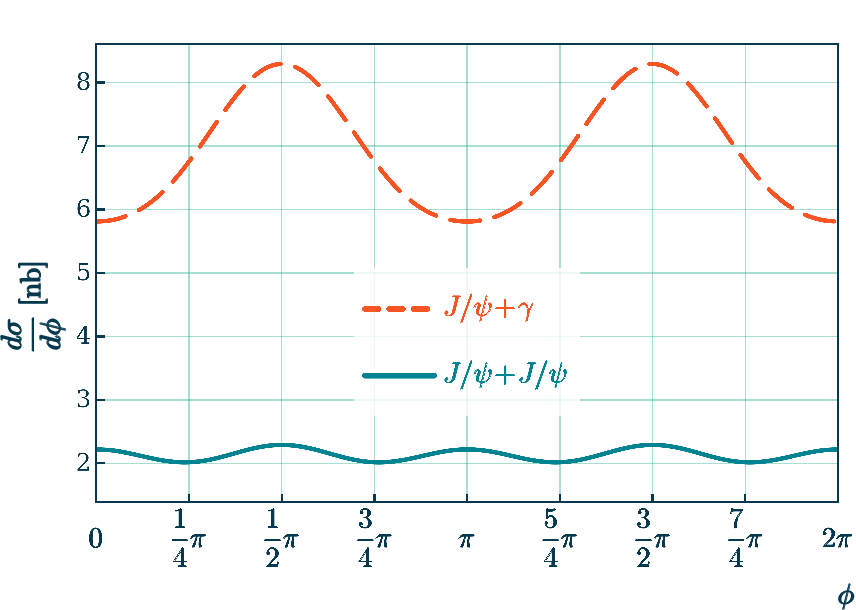}
\caption{The azimuthal distribution $d\sigma / d\phi$ for $J/\psi + \gamma(J/\psi)$ production in Pb+Pb UPCs 
at $\sqrt{s_{\rm NN}} = 5.02$~TeV, integrated over $Q$ from threshold to 20 GeV and $q_\perp$ from 0 to 200 MeV, with cuts $1.6 < |y_{1,2}| < 2.4$ and $P_\perp > 200$~MeV.}
\label{fig:ds_dphi}
\end{center}
\end{figure}

\begin{figure}[htbp]
\begin{center}
\includegraphics[width=0.425\textwidth]{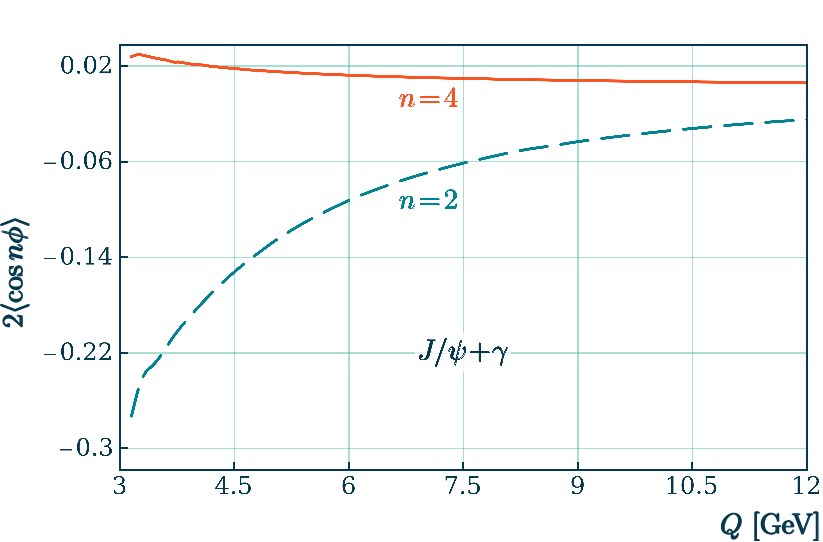}
\includegraphics[width=0.425\textwidth]{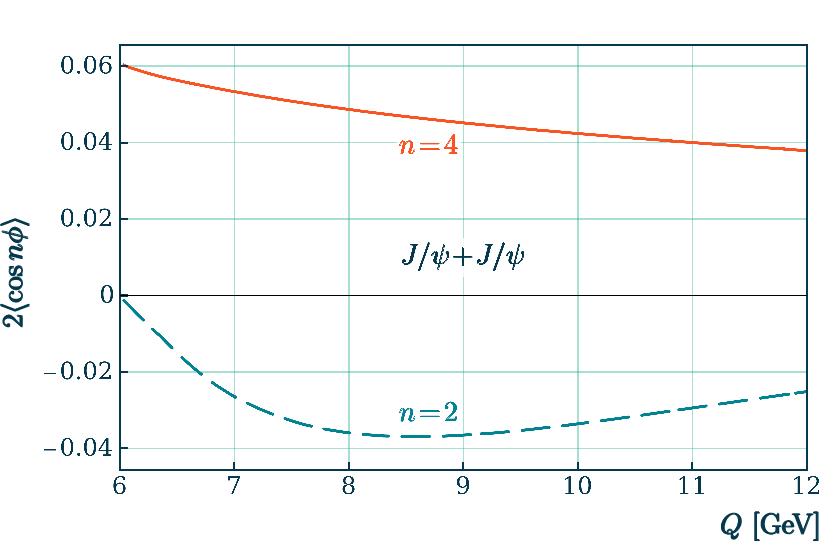}
\caption{The azimuthal asymmetry observables $\langle \cos(2\phi) \rangle$ and $\langle \cos(4\phi) \rangle$, 
as a function of the invariant mass $Q$ for $J/\psi + \gamma$ and $J/\psi + J/\psi$ production in Pb+Pb UPCs at 
$\sqrt{s_{\rm NN}} = 5.02$~TeV, under the kinematic constraints $1.6 < |y_{1,2}| < 2.4$, $Q < 20$~GeV and $P_\perp > 200$~MeV.}
\label{fig:asym_Q}
\end{center}
\end{figure}

We plot the azimuthal distribution $d\sigma/d\phi$ in Fig.~\ref{fig:ds_dphi}. 
The $\cos(2\phi)$ and $\cos(4\phi)$ modulation behaviors are visible for both processes, highlighting the crucial impact of 
the photon linear polarization in generating these angular correlations. 
These azimuthal asymmetries can be quantitatively visualized in Fig.~\ref{fig:asym_Q}. 
For both processes $\langle \cos(2\phi) \rangle$ is negative while $\langle \cos(4\phi) \rangle$ is positive. 
For $J/\psi + \gamma$ production, $\langle \cos(2\phi) \rangle$ dominates the asymmetry, 
with its absolute value exceeding $13\%$ near the production threshold.  
For $J/\psi + J/\psi$ production, while $\langle \cos(4\phi) \rangle$ has a larger absolute value than $\langle \cos(2\phi) \rangle$, 
the maximal asymmetry reaches only about $3\%$ near the threshold.

One may wonder how these predicted azimuthal asymmetries are influenced by radiative and relativistic corrections. Fortunately, as ratios of cross-sections, the azimuthal observables defined in 
Eq.~\ref{azimuthal:variable} are expected to remain stable when higher-order corrections are included. For the same reason, we anticipate that our predictions are only mildly sensitive to the precise values of the NRQCD LDMEs, which currently carry significant uncertainties.
\section{Conclusion}

In this paper, we have presented the first TMD factorization analysis of $J/\psi+\gamma$ and $J/\psi+J/\psi$ 
production in ultraperipheral Pb-Pb collisions at {\tt LHC}. 
Taking the LO NRQCD predictions for the helicity amplitudes of $\gamma\gamma\to J/\psi+\gamma (J/\psi)$ as input,
we have demonstrated that the linear polarization of quasi-real photons do induce azimuthal correlation in the final state.

Our central finding is the prediction of sizable $\cos(2\phi)$ and $\cos(4\phi)$ asymmetries, 
reaching up to $13\%$ in accessible kinematic regions at the LHC. 
These modulations arise purely from the quantum interference between distinct photon helicity states—a phenomenon completely 
missed by the familiar EPA approach. The cleanliness of the UPC environment together with the clearness of the $J/\psi$ and
photon signals,  ensures that these azimuthal observables can serve a 
novel testing ground for exclusive quarkonium production mechanism and the photon TMD structure in the ultrarelativistic limit.

As the observables built out of ratios of weighted to unweighted cross sections, these azimuthal asymmetries are expected to be 
largely immune to systematic uncertainties associated with integrated luminosity, 
the absolute normalization of NRQCD LDMEs as well as higher order corrections. 
Concerning the rich spin information encoded in the 
$\gamma\gamma\to J/\psi+\gamma(J/\psi)$ amplitudes, it appears interesting to 
further investigate the azimuthal correlation for 
{\it polarized} $J/\psi$ production in UPCs, 
as well as explore the quantum entanglement and  spin correlation 
between two outgoing polarized $J/\psi$. 

\

\begin{acknowledgments}
We are grateful to Shuai Yang for helpful discussions. This work was supported by the High Performance Computing Center of Central South University.
The work of Y.~J. is supported in part by the NNSFC under Grant No.~12475090.
The work of W.~S. is supported in part by the NNSFC under Grant No.~12375079.
The work of X.~X. is supported in part by the NNSFC under Grant No.~12275364.
The work of J.~Z. is supported in part by the NNSFC under Grant No. 12175118 and No.~12321005.
The work of Y.~Z. is supported in part by the NNSFC (Grants No. 12475084) and the Natural Science Foundation of Shandong Province (Grant No. ZR2024MA012).
\end{acknowledgments}


\newpage

\appendix

\begin{widetext}
 
\section{Squared Helicity Amplitudes for $\gamma\gamma \to J/\psi + \gamma (J/\psi)$}

In this appendix, we present explicit expressions for the interference of 
one helicity amplitude with another conjugated one, {\it viz.},   $M_{\lambda_{1},\lambda_{2}}M_{\lambda_{1}^{\prime},\lambda_{2}^{\prime}}^{*}$
in Eq.~\eqref{semi-inclusive:production:cross:section}. 
The helicity amplitudes are projected from the LO NRQCD predictions for 
the $\gamma\gamma \to J/\psi + \gamma (J/\psi)$ amplitudes. 
These squared helicity amplitudes depend upon the Mandelstam variables $s$, $t$, $u$ and charm quark 
mass $m_c$, as well as upon the lowest-order NRQCD LDME of $J/\psi$, often approximated 
by the wave function at the origin for $J/\psi$. 

For the $\gamma\gamma\to J/\psi+\gamma$ reaction, the three independent squared helicity amplitudes read
\bseq
\bqa
|M_{++}|^{2}&=&\mathcal{N}_1 \frac{s^2}{(s+t)^2(s+u)^2}, \\
|M_{+-}|^{2}&=&\mathcal{N}_1 \frac{s^2(t^2+u^2)+2 s t u (t+u)+2 t^2 u^2}{(s+t)^2(s+u)^2(t+u)^2}, \\
M_{++}M_{+-}^{*}&=&M_{+-} M_{++}^{*}=\mathcal{N}_1 \frac{4 m_c^2 s\, t\, u}{(s+t)^2(s+u)^2(t+u)^2},
\eqa
\eseq
where the normalization factor $\mathcal{N}_1 = 24576\, \pi^2 \alpha^3 e_c^6 m_c |R_{J/\psi}(0)|^2$, with $\alpha$ the QED coupling constant and $e_c=2/3$.

For the $\gamma\gamma\to J/\psi+J/\psi$ channel, the results become more involved relative to the $J/\psi+\gamma$ production, due to the presence of the extra gluon exchange. The three independent squared helicity amplitudes read
\bseq
\bqa
|M_{++}|^{2}&=&\mathcal{N}_2 (s-12 m_c^2) (s-4 m_c^2)
   \bigg[\frac{s^3}{(t-4 m_c^2)^4}+\frac{2
   s}{(t-4 m_c^2)^2}-\frac{4}{t-4 m_c^2}+\bigg(t \leftrightarrow u\bigg)\bigg], \\
|M_{+-}|^{2}&=&\mathcal{N}_2 \bigg[\frac{16 m_c^2 s^3}{(t-4 m_c^2)^3}+\frac{s^3
   (4 m_c^2+s) (12 m_c^2+s)}{(t-4 m_c^2)^4}+\frac{2 s
   (4 m_c^2+s) (12 m_c^2+5
   s)}{(t-4 m_c^2)^2}\nn\\
   &&+\frac{1}{32} s
   \left(\frac{s^2}{m_c^4}-\frac{16
   s}{m_c^2}+192\right)-\frac{192 m_c^6+32 m_c^4
   s+s^3}{m_c^2 (t-4 m_c^2)}+\bigg(t \leftrightarrow u\bigg)\bigg], \\
M_{++}M_{+-}^{*}&=&M_{+-} M_{++}^{*}=\mathcal{N}_2 \bigg[\frac{s (-96
   m_c^4-16 m_c^2 s+s^2)}{(t-4
   m_c^2)^2}-\frac{8 m_c^2 s^3}{(t-4 m_c^2)^3}-\frac{16 m_c^2 (s-12 m_c^2)}{t-4 m_c^2}-\frac{48
   m_c^4 s^3}{(t-4 m_c^2)^4}\nn\\
   && +\bigg(t \leftrightarrow u\bigg) \bigg],
\eqa
\eseq
 where the normalization factor 
$\mathcal{N}_2 = \frac{4194304\, \pi^2\alpha_s^2 \alpha^2 e_c^4 m_c^2 |R_{J/\psi}(0)|^4}{9 s^5}$.

\end{widetext}
 
\bibliography{ref}

@article{Ma:2015vpt,
    author = "Ma, J. P. and Wang, C.",
    title = "{QCD factorization for quarkonium production in hadron collisions at low transverse momentum}",
    eprint = "1509.04421",
    archivePrefix = "arXiv",
    primaryClass = "hep-ph",
    doi = "10.1103/PhysRevD.93.014025",
    journal = "Phys. Rev. D",
    volume = "93",
    number = "1",
    pages = "014025",
    year = "2016"
}

@article{Taels:2022tza,
    author = "Taels, Pieter and Altinoluk, Tolga and Beuf, Guillaume and Marquet, Cyrille",
    title = "{Dijet photoproduction at low x at next-to-leading order and its back-to-back limit}",
    eprint = "2204.11650",
    archivePrefix = "arXiv",
    primaryClass = "hep-ph",
    doi = "10.1007/JHEP10(2022)184",
    journal = "JHEP",
    volume = "10",
    pages = "184",
    year = "2022"
}

@article{Caucal:2023fsf,
    author = {Caucal, Paul and Salazar, Farid and Schenke, Bj{\"o}rn and Stebel, Tomasz and Venugopalan, Raju},
    title = "{Back-to-Back Inclusive Dijets in Deep Inelastic Scattering at Small x: Complete NLO Results and Predictions}",
    eprint = "2308.00022",
    archivePrefix = "arXiv",
    primaryClass = "hep-ph",
    doi = "10.1103/PhysRevLett.132.081902",
    journal = "Phys. Rev. Lett.",
    volume = "132",
    number = "8",
    pages = "081902",
    year = "2024"
}

@article{Maxia:2024cjh,
    author = "Maxia, Luca and Yuan, Feng",
    title = "{Azimuthal angular correlation of J/{\ensuremath{\psi}} plus jet production at the electron-ion collider}",
    eprint = "2403.02097",
    archivePrefix = "arXiv",
    primaryClass = "hep-ph",
    doi = "10.1103/PhysRevD.110.114042",
    journal = "Phys. Rev. D",
    volume = "110",
    number = "11",
    pages = "114042",
    year = "2024"
}

@article{Bodwin:1994jh,
    author = "Bodwin, Geoffrey T. and Braaten, Eric and Lepage, G. Peter",
    title = "{Rigorous QCD analysis of inclusive annihilation and production of heavy quarkonium}",
    eprint = "hep-ph/9407339",
    archivePrefix = "arXiv",
    reportNumber = "ANL-HEP-PR-94-24, FERMILAB-PUB-94-073-T, NUHEP-TH-94-5",
    doi = "10.1103/PhysRevD.55.5853",
    journal = "Phys. Rev. D",
    volume = "51",
    pages = "1125--1171",
    year = "1995",
    note = "[Erratum: Phys.Rev.D 55, 5853 (1997)]"
}

@inproceedings{Haber:1994pe,
    author = "Haber, Howard E.",
    title = "{Spin formalism and applications to new physics searches}",
    booktitle = "{21st Annual SLAC Summer Institute on Particle Physics: Spin Structure in High-energy Processes (School: 26 Jul - 3 Aug, Topical Conference: 4-6 Aug) (SSI 93)}",
    eprint = "hep-ph/9405376",
    archivePrefix = "arXiv",
    reportNumber = "SCIPP-93-49, NSF-ITP-94-30",
    pages = "231--272",
    month = "4",
    year = "1994"
}

@article{denDunnen:2014kjo,
    author = "den Dunnen, Wilco J. and Lansberg, J. P. and Pisano, C. and Schlegel, M.",
    title = "{Accessing the Transverse Dynamics and Polarization of Gluons inside the Proton at the LHC}",
    eprint = "1401.7611",
    archivePrefix = "arXiv",
    primaryClass = "hep-ph",
    reportNumber = "NIKHEF-2013-040",
    doi = "10.1103/PhysRevLett.112.212001",
    journal = "Phys. Rev. Lett.",
    volume = "112",
    pages = "212001",
    year = "2014"
}

@article{Alimov:2024pqt,
    author = "Alimov, Lev and Karpishkov, Anton and Saleev, Vladimir",
    title = "{Associated production of J{\ensuremath{/}}{\ensuremath{\psi}} and direct photon in the NRQCD and the ICEM using the high-energy factorization}",
    eprint = "2412.01710",
    archivePrefix = "arXiv",
    primaryClass = "hep-ph",
    doi = "10.1142/S0217751X25500204",
    journal = "Int. J. Mod. Phys. A",
    volume = "40",
    number = "15",
    pages = "2550020",
    year = "2025"
}

@article{liu:2023unl,
    author = "liu, Huachao and xie, Xiupeng and Lu, Zhun",
    title = "{Gluon TMDs from J/{\ensuremath{\psi}} production in longitudinally polarized deeply inelastic scattering}",
    eprint = "2310.00609",
    archivePrefix = "arXiv",
    primaryClass = "hep-ph",
    doi = "10.1016/j.physletb.2023.138439",
    journal = "Phys. Lett. B",
    volume = "849",
    pages = "138439",
    year = "2024"
}

@article{Kishore:2021vsm,
    author = "Kishore, Raj and Mukherjee, Asmita and Siddiqah, Mariyah",
    title = "{Cos(2$\phi_h$) asymmetry in J/$\psi$ production in unpolarized $ep$ collision}",
    eprint = "2103.09070",
    archivePrefix = "arXiv",
    primaryClass = "hep-ph",
    doi = "10.1103/PhysRevD.104.094015",
    journal = "Phys. Rev. D",
    volume = "104",
    number = "9",
    pages = "094015",
    year = "2021"
}

@article{Boer:2020bbd,
    author = {Boer, Dani{\"e}l and D'Alesio, Umberto and Murgia, Francesco and Pisano, Cristian and Taels, Pieter},
    title = "{J/{\ensuremath{\psi}} meson production in SIDIS: matching high and low transverse momentum}",
    eprint = "2004.06740",
    archivePrefix = "arXiv",
    primaryClass = "hep-ph",
    doi = "10.1007/JHEP09(2020)040",
    journal = "JHEP",
    volume = "09",
    pages = "040",
    year = "2020"
}

@article{Scarpa:2019fol,
    author = {Scarpa, Florent and Boer, Dani{\"e}l and Echevarria, Miguel G. and Lansberg, Jean-Philippe and Pisano, Cristian and Schlegel, Marc},
    title = "{Studies of gluon TMDs and their evolution using quarkonium-pair production at the LHC}",
    eprint = "1909.05769",
    archivePrefix = "arXiv",
    primaryClass = "hep-ph",
    doi = "10.1140/epjc/s10052-020-7619-1",
    journal = "Eur. Phys. J. C",
    volume = "80",
    number = "2",
    pages = "87",
    year = "2020"
}

@article{Bacchetta:2018ivt,
    author = {Bacchetta, Alessandro and Boer, Dani{\"e}l and Pisano, Cristian and Taels, Pieter},
    title = "{Gluon TMDs and NRQCD matrix elements in $J/\psi$ production at an EIC}",
    eprint = "1809.02056",
    archivePrefix = "arXiv",
    primaryClass = "hep-ph",
    doi = "10.1140/epjc/s10052-020-7620-8",
    journal = "Eur. Phys. J. C",
    volume = "80",
    number = "1",
    pages = "72",
    year = "2020"
}

@article{Lansberg:2017dzg,
    author = "Lansberg, Jean-Philippe and Pisano, Cristian and Scarpa, Florent and Schlegel, Marc",
    title = "{Pinning down the linearly-polarised gluons inside unpolarised protons using quarkonium-pair production at the LHC}",
    eprint = "1710.01684",
    archivePrefix = "arXiv",
    primaryClass = "hep-ph",
    doi = "10.1016/j.physletb.2018.08.004",
    journal = "Phys. Lett. B",
    volume = "784",
    pages = "217--222",
    year = "2018",
    note = "[Erratum: Phys.Lett.B 791, 420--421 (2019)]"
}

@article{Zhang:2017uiz,
    author = "Zhang, Guang-Peng",
    title = "{Back-to-back heavy quark pair production in Semi-inclusive DIS}",
    eprint = "1709.08970",
    archivePrefix = "arXiv",
    primaryClass = "hep-ph",
    doi = "10.1007/JHEP11(2017)069",
    journal = "JHEP",
    volume = "11",
    pages = "069",
    year = "2017"
}

@article{Lansberg:2016abn,
    author = "Lansberg, Jean-Philippe and Shao, Hua-Sheng",
    editor = "Foka, Y. and Brambilla, N. and Kovalenko, V.",
    title = "{$J/\psi + Z$ production at the LHC}",
    eprint = "1611.10306",
    archivePrefix = "arXiv",
    primaryClass = "hep-ph",
    doi = "10.1051/epjconf/201713706013",
    journal = "EPJ Web Conf.",
    volume = "137",
    pages = "06013",
    year = "2017"
}

@article{Boer:2016fqd,
    author = {Boer, Dani{\"e}l and Mulders, Piet J. and Pisano, Cristian and Zhou, Jian},
    title = "{Asymmetries in Heavy Quark Pair and Dijet Production at an EIC}",
    eprint = "1605.07934",
    archivePrefix = "arXiv",
    primaryClass = "hep-ph",
    doi = "10.1007/JHEP08(2016)001",
    journal = "JHEP",
    volume = "08",
    pages = "001",
    year = "2016"
}

@article{Boer:2017xpy,
    author = "Boer, Daniel and Mulders, Piet J. and Zhou, Jian and Zhou, Ya-jin",
    title = "{Suppression of maximal linear gluon polarization in angular asymmetries}",
    eprint = "1702.08195",
    archivePrefix = "arXiv",
    primaryClass = "hep-ph",
    doi = "10.1007/JHEP10(2017)196",
    journal = "JHEP",
    volume = "10",
    pages = "196",
    year = "2017"
}

@article{Zhang:2014vmh,
    author = "Zhang, Guang-Peng",
    title = "{Probing transverse momentum dependent gluon distribution functions from hadronic quarkonium pair production}",
    eprint = "1406.5476",
    archivePrefix = "arXiv",
    primaryClass = "hep-ph",
    doi = "10.1103/PhysRevD.90.094011",
    journal = "Phys. Rev. D",
    volume = "90",
    number = "9",
    pages = "094011",
    year = "2014"
}

@article{Ma:2014oha,
    author = "Ma, J. P. and Wang, J. X. and Zhao, S.",
    title = "{Breakdown of QCD Factorization for P-Wave Quarkonium Production at Low Transverse Momentum}",
    eprint = "1405.3373",
    archivePrefix = "arXiv",
    primaryClass = "hep-ph",
    doi = "10.1016/j.physletb.2014.08.033",
    journal = "Phys. Lett. B",
    volume = "737",
    pages = "103--108",
    year = "2014"
}

@article{Chen:2025gwp,
    author = "Chen, Zi-Qiang and Chen, Long-Bin",
    title = "{Exclusive J/{\ensuremath{\psi}}+{\ensuremath{\gamma}} production in ultraperipheral ion collisions}",
    eprint = "2508.06946",
    archivePrefix = "arXiv",
    primaryClass = "hep-ph",
    doi = "10.1103/p3yc-hqly",
    journal = "Phys. Rev. D",
    volume = "112",
    number = "5",
    pages = "056013",
    year = "2025"
}

@article{Bodwin:2007ga,
    author = "Bodwin, Geoffrey T. and Lee, Jungil and Yu, Chaehyun",
    title = "{Resummation of Relativistic Corrections to e+ e- ---{\ensuremath{>}} J/psi + eta(c)}",
    eprint = "0710.0995",
    archivePrefix = "arXiv",
    primaryClass = "hep-ph",
    reportNumber = "ANL-HEP-PR-07-79",
    doi = "10.1103/PhysRevD.77.094018",
    journal = "Phys. Rev. D",
    volume = "77",
    pages = "094018",
    year = "2008"
}

@article{Hagiwara:2017fye,
    author = "Hagiwara, Yoshikazu and Hatta, Yoshitaka and Pasechnik, Roman and Tasevsky, Marek and Teryaev, Oleg",
    title = "{Accessing the gluon Wigner distribution in ultraperipheral $pA$ collisions}",
    eprint = "1706.01765",
    archivePrefix = "arXiv",
    primaryClass = "hep-ph",
    reportNumber = "YITP-17-56, LU-TP-17-25",
    doi = "10.1103/PhysRevD.96.034009",
    journal = "Phys. Rev. D",
    volume = "96",
    number = "3",
    pages = "034009",
    year = "2017"
}

@article{Li:2019yzy,
    author = "Li, Cong and Zhou, Jian and Zhou, Ya-Jin",
    title = "{Probing the linear polarization of photons in ultraperipheral heavy ion collisions}",
    eprint = "1903.10084",
    archivePrefix = "arXiv",
    primaryClass = "hep-ph",
    doi = "10.1016/j.physletb.2019.07.005",
    journal = "Phys. Lett. B",
    volume = "795",
    pages = "576--580",
    year = "2019"
}

@article{Li:2019sin,
    author = "Li, Cong and Zhou, Jian and Zhou, Ya-Jin",
    title = "{Impact parameter dependence of the azimuthal asymmetry in lepton pair production in heavy ion collisions}",
    eprint = "1911.00237",
    archivePrefix = "arXiv",
    primaryClass = "hep-ph",
    doi = "10.1103/PhysRevD.101.034015",
    journal = "Phys. Rev. D",
    volume = "101",
    number = "3",
    pages = "034015",
    year = "2020"
}

@article{Vidovic:1992ik,
    author = "Vidovic, M. and Greiner, M. and Best, C. and Soff, G.",
    title = "{Impact parameter dependence of the electromagnetic particle production in ultrarelativistic heavy ion collisions}",
    reportNumber = "GSI-92-52",
    doi = "10.1103/PhysRevC.47.2308",
    journal = "Phys. Rev. C",
    volume = "47",
    pages = "2308--2319",
    year = "1993"
}

@article{Klein:2018fmp,
    author = "Klein, Spencer and Mueller, A.H. and Xiao, Bo-Wen and Yuan, Feng",
    title = "{Acoplanarity of a Lepton Pair to Probe the Electromagnetic Property of Quark Matter}",
    eprint = "1811.05519",
    archivePrefix = "arXiv",
    primaryClass = "hep-ph",
    doi = "10.1103/PhysRevLett.122.132301",
    journal = "Phys. Rev. Lett.",
    volume = "122",
    number = "13",
    pages = "132301",
    year = "2019"
}

@article{Zha:2018tlq,
    author = "Zha, Wangmei and Brandenburg, James Daniel and Tang, Zebo and Xu, Zhangbu",
    title = "{Initial transverse-momentum broadening of Breit-Wheeler process in relativistic heavy-ion collisions}",
    eprint = "1812.02820",
    archivePrefix = "arXiv",
    primaryClass = "nucl-th",
    doi = "10.1016/j.physletb.2019.135089",
    journal = "Phys. Lett. B",
    volume = "800",
    pages = "135089",
    year = "2020"
}

@article{Klein:2016yzr,
    author = "Klein, Spencer R. and Nystrand, Joakim and Seger, Janet and Gorbunov, Yuri and Butterworth, Joey",
    title = "{STARlight: A Monte Carlo simulation program for ultra-peripheral collisions of relativistic ions}",
    eprint = "1607.03838",
    archivePrefix = "arXiv",
    primaryClass = "hep-ph",
    doi = "10.1016/j.cpc.2016.10.016",
    journal = "Comput. Phys. Commun.",
    volume = "212",
    pages = "258--268",
    year = "2017"
}

@article{Hencken:1994my,
    author = "Hencken, Kai and Trautmann, Dirk and Baur, Gerhard",
    title = "{Impact parameter dependence of the total probability for the electromagnetic electron - positron pair production in relativistic heavy ion collisions}",
    eprint = "nucl-th/9410014",
    archivePrefix = "arXiv",
    reportNumber = "KFA-IKP-TH-1994-38",
    doi = "10.1103/PhysRevA.51.1874",
    journal = "Phys. Rev. A",
    volume = "51",
    pages = "1874--1882",
    year = "1995"
}

@article{Klein:2020jom,
    author = "Klein, Spencer and Mueller, A.H. and Xiao, Bo-Wen and Yuan, Feng",
    title = "{Lepton Pair Production Through Two Photon Process in Heavy Ion Collisions}",
    eprint = "2003.02947",
    archivePrefix = "arXiv",
    primaryClass = "hep-ph",
    month = "3",
    year = "2020"
}

@article{Xing:2020hwh,
    author = "Xing, Hongxi and Zhang, Cheng and Zhou, Jian and Zhou, Ya-Jin",
    title = "{The cos 2$\phi$ azimuthal asymmetry in $\rho^{0}$ meson production in ultraperipheral heavy ion collisions}",
    eprint = "2006.06206",
    archivePrefix = "arXiv",
    primaryClass = "hep-ph",
    doi = "10.1007/JHEP10(2020)064",
    journal = "JHEP",
    volume = "10",
    pages = "064",
    year = "2020"
}

@article{Xiao:2020ddm,
    author = "Xiao, Bo-Wen and Yuan, Feng and Zhou, Jian",
    title = "{Electromagnetic Flow of Leptons in Heavy Ion Collisions}",
    eprint = "2003.06352",
    archivePrefix = "arXiv",
    primaryClass = "hep-ph",
    month = "3",
    year = "2020"
}

@article{Metz:2011wb,
    author = "Metz, Andreas and Zhou, Jian",
    title = "{Distribution of linearly polarized gluons inside a large nucleus}",
    eprint = "1105.1991",
    archivePrefix = "arXiv",
    primaryClass = "hep-ph",
    doi = "10.1103/PhysRevD.84.051503",
    journal = "Phys. Rev. D",
    volume = "84",
    pages = "051503",
    year = "2011"
}

@Article{Wang:2022gkd,
  author        = {Wang, Ren-jie and Lin, Shuo and Pu, Shi and Zhang, Yi-fei and Wang, Qun},
  journal       = {Phys. Rev. D},
  title         = {{Lepton pair photoproduction in peripheral relativistic heavy-ion collisions}},
  year          = {2022},
  number        = {3},
  pages         = {034025},
  volume        = {106},
  archiveprefix = {arXiv},
  doi           = {10.1103/PhysRevD.106.034025},
  eprint        = {2204.02761},
  primaryclass  = {hep-ph},
}

@Article{Wang:2021kxm,
  author        = {Wang, Ren-jie and Pu, Shi and Wang, Qun},
  journal       = {Phys. Rev. D},
  title         = {{Lepton pair production in ultraperipheral collisions}},
  year          = {2021},
  number        = {5},
  pages         = {056011},
  volume        = {104},
  archiveprefix = {arXiv},
  doi           = {10.1103/PhysRevD.104.056011},
  eprint        = {2106.05462},
  primaryclass  = {hep-ph},
  reportnumber  = {USTC-ICTS/PCFT-21-25},
}

@Article{Shao:2024nor,
  author        = {Shao, Ding Yu and Shi, Yu and Zhang, Cheng and Zhou, Jian and Zhou, Ya-jin},
  journal       = {JHEP},
  title         = {{Revisiting azimuthal angular asymmetries in diffractive di-jet production}},
  year          = {2024},
  pages         = {189},
  volume        = {07},
  archiveprefix = {arXiv},
  doi           = {10.1007/JHEP07(2024)189},
  eprint        = {2402.05465},
  primaryclass  = {hep-ph},
}

@Article{Iancu:2023lel,
  author        = {Iancu, E. and Mueller, A. H. and Triantafyllopoulos, D. N. and Wei, S. Y.},
  journal       = {Eur. Phys. J. C},
  title         = {{Probing gluon saturation via diffractive jets in ultra-peripheral nucleus-nucleus collisions}},
  year          = {2023},
  number        = {11},
  pages         = {1078},
  volume        = {83},
  archiveprefix = {arXiv},
  doi           = {10.1140/epjc/s10052-023-12165-8},
  eprint        = {2304.12401},
  primaryclass  = {hep-ph},
}

@Article{AH:2023kor,
  author        = {A H, Ajjath and Chaubey, Ekta and Fraaije, Mathijs and Hirschi, Valentin and Shao, Hua-Sheng},
  journal       = {Phys. Lett. B},
  year          = {2024},
  pages         = {138555},
  volume        = {851},
  archiveprefix = {arXiv},
  doi           = {10.1016/j.physletb.2024.138555},
  eprint        = {2312.16956},
  primaryclass  = {hep-ph},
}

@Article{Shao:2023zge,
  author        = {Shao, Ding Yu and Zhang, Cheng and Zhou, Jian and Zhou, Ya-jin},
  journal       = {Phys. Rev. D},
  title         = {{Lepton pair production in ultraperipheral collisions: Toward a precision test of the resummation formalism}},
  year          = {2023},
  number        = {11},
  pages         = {116015},
  volume        = {108},
  archiveprefix = {arXiv},
  doi           = {10.1103/PhysRevD.108.116015},
  eprint        = {2306.02337},
  primaryclass  = {hep-ph},
}

@Article{Harland-Lang:2020veo,
  author        = {Harland-Lang, L. A. and Tasevsky, M. and Khoze, V. A. and Ryskin, M. G.},
  journal       = {Eur. Phys. J. C},
  title         = {{A new approach to modelling elastic and inelastic photon-initiated production at the LHC: SuperChic 4}},
  year          = {2020},
  number        = {10},
  pages         = {925},
  volume        = {80},
  archiveprefix = {arXiv},
  doi           = {10.1140/epjc/s10052-020-08455-0},
  eprint        = {2007.12704},
  primaryclass  = {hep-ph},
  reportnumber  = {IPPP/20/33},
}

@Article{Shao:2022cly,
  author        = {Shao, Hua-Sheng and d'Enterria, David},
  journal       = {JHEP},
  title         = {{gamma-UPC: automated generation of exclusive photon-photon processes in ultraperipheral proton and nuclear collisions with varying form factors}},
  year          = {2022},
  pages         = {248},
  volume        = {09},
  archiveprefix = {arXiv},
  doi           = {10.1007/JHEP09(2022)248},
  eprint        = {2207.03012},
  primaryclass  = {hep-ph},
}

@Article{Mantysaari:2023prg,
  author        = {M\"antysaari, Heikki and Salazar, Farid and Schenke, Bj\"orn and Shen, Chun and Zhao, Wenbin},
  journal       = {Phys. Rev. C},
  title         = {{Effects of nuclear structure and quantum interference on diffractive vector meson production in ultraperipheral nuclear collisions}},
  year          = {2024},
  number        = {2},
  pages         = {024908},
  volume        = {109},
  archiveprefix = {arXiv},
  doi           = {10.1103/PhysRevC.109.024908},
  eprint        = {2310.15300},
  primaryclass  = {nucl-th},
}

@Article{Hagiwara:2020juc,
  author        = {Hagiwara, Yoshikazu and Zhang, Cheng and Zhou, Jian and Zhou, Ya-Jin},
  journal       = {Phys. Rev. D},
  title         = {{Coulomb nuclear interference effect in dipion production in ultraperipheral heavy ion collisions}},
  year          = {2021},
  number        = {7},
  pages         = {074013},
  volume        = {103},
  archiveprefix = {arXiv},
  doi           = {10.1103/PhysRevD.103.074013},
  eprint        = {2011.13151},
  primaryclass  = {hep-ph},
}

@Article{Hagiwara:2021xkf,
  author        = {Hagiwara, Yoshikazu and Zhang, Cheng and Zhou, Jian and Zhou, Ya-jin},
  journal       = {Phys. Rev. D},
  title         = {{Probing the gluon tomography in photoproduction of dipion}},
  year          = {2021},
  number        = {9},
  pages         = {094021},
  volume        = {104},
  archiveprefix = {arXiv},
  doi           = {10.1103/PhysRevD.104.094021},
  eprint        = {2106.13466},
  primaryclass  = {hep-ph},
}

@Article{Zhang:2024mql,
  author        = {Zhang, Cheng and Zhang, Li-Mao and Shao, Ding Yu},
  title         = {{Photon induced proton and anti-proton pair production with ultraperipheral heavy ion collisions at RHIC}},
  year          = {2024},
  month         = {6},
  archiveprefix = {arXiv},
  eprint        = {2406.05618},
  primaryclass  = {hep-ph},
}

@Article{Lin:2024mnj,
  author        = {Lin, Shuo and Hu, Jin-Yu and Xu, Hao-Jie and Pu, Shi and Wang, Qun},
  title         = {{Nuclear deformation effects in photoproduction of $\rho$ mesons in ultraperipheral isobaric collisions}},
  year          = {2024},
  month         = {5},
  archiveprefix = {arXiv},
  eprint        = {2405.16491},
  primaryclass  = {hep-ph},
}

@Article{Hu:2024bsm,
  author        = {Hu, Jin-Yu and Lin, Shuo and Pu, Shi and Wang, Qun},
  title         = {{Light nuclei photoproduction in relativistic heavy ion ultraperipheral collisions}},
  year          = {2024},
  month         = {7},
  archiveprefix = {arXiv},
  eprint        = {2407.06091},
  primaryclass  = {hep-ph},
}

@Article{ALICE:2020ugp,
  author        = {Acharya, Shreyasi and others},
  journal       = {JHEP},
  title         = {{Coherent photoproduction of $\rho^{0}$ vector mesons in ultra-peripheral Pb-Pb collisions at $ \sqrt{{\mathrm{s}}_{\mathrm{NN}}} $ = 5.02 TeV}},
  year          = {2020},
  pages         = {035},
  volume        = {06},
  archiveprefix = {arXiv},
  collaboration = {ALICE},
  doi           = {10.1007/JHEP06(2020)035},
  eprint        = {2002.10897},
  primaryclass  = {nucl-ex},
  reportnumber  = {CERN-EP-2020-021},
}

@Article{ALICE:2021gpt,
  author        = {Acharya, Shreyasi and others},
  journal       = {Eur. Phys. J. C},
  title         = {{Coherent $J/\psi$ and $\psi'$ photoproduction at midrapidity in ultra-peripheral Pb-Pb collisions at $\sqrt{s_{\mathrm{NN}}}~=~5.02$ TeV}},
  year          = {2021},
  number        = {8},
  pages         = {712},
  volume        = {81},
  archiveprefix = {arXiv},
  collaboration = {ALICE},
  doi           = {10.1140/epjc/s10052-021-09437-6},
  eprint        = {2101.04577},
  primaryclass  = {nucl-ex},
  reportnumber  = {CERN-EP-2021-002},
}

@Article{Brandenburg:2024ksp,
  author        = {Brandenburg, James Daniel and Duan, Haowu and Tu, Zhoudunming and Venugopalan, Raju and Xu, Zhangbu},
  title         = {{Entanglement Enabled Intensity Interferometry in ultrarelativistic ultraperipheral nuclear collisions}},
  year          = {2024},
  month         = {7},
  archiveprefix = {arXiv},
  eprint        = {2407.15945},
  primaryclass  = {hep-ph},
}

@article{ALICE:2019tqa,
    author = "Acharya, Shreyasi and others",
    collaboration = "ALICE",
    title = "{Coherent J/$\psi$ photoproduction at forward rapidity in ultra-peripheral Pb-Pb collisions at $\sqrt{s_{\rm{NN}}}=5.02$ TeV}",
    eprint = "1904.06272",
    archivePrefix = "arXiv",
    primaryClass = "nucl-ex",
    reportNumber = "CERN-EP-2019-069",
    doi = "10.1016/j.physletb.2019.134926",
    journal = "Phys. Lett. B",
    volume = "798",
    pages = "134926",
    year = "2019"
}

@Article{STAR:2019wlg,
  author        = {Adam, Jaroslav and others},
  journal       = {Phys. Rev. Lett.},
  title         = {{Measurement of $e^+e^-$ Momentum and Angular Distributions from Linearly Polarized Photon Collisions}},
  year          = {2021},
  number        = {5},
  pages         = {052302},
  volume        = {127},
  archiveprefix = {arXiv},
  collaboration = {STAR},
  doi           = {10.1103/PhysRevLett.127.052302},
  eprint        = {1910.12400},
  primaryclass  = {nucl-ex},
}

@Article{STAR:2022wfe,
  author        = {Abdallah, Mohamed and others},
  journal       = {Sci. Adv.},
  title         = {{Tomography of ultrarelativistic nuclei with polarized photon-gluon collisions}},
  year          = {2023},
  number        = {1},
  pages         = {eabq3903},
  volume        = {9},
  archiveprefix = {arXiv},
  collaboration = {STAR},
  doi           = {10.1126/sciadv.abq3903},
  eprint        = {2204.01625},
  primaryclass  = {nucl-ex},
}

@article{Bertulani:1987tz,
    author = "Bertulani, Carlos A. and Baur, Gerhard",
    title = "{Electromagnetic Processes in Relativistic Heavy Ion Collisions}",
    reportNumber = "JUL-2163",
    doi = "10.1016/0370-1573(88)90142-1",
    journal = "Phys. Rept.",
    volume = "163",
    pages = "299",
    year = "1988"
}

@Article{Bertulani:2005ru,
  author        = {Bertulani, Carlos A. and Klein, Spencer R. and Nystrand, Joakim},
  journal       = {Ann. Rev. Nucl. Part. Sci.},
  title         = {{Physics of ultra-peripheral nuclear collisions}},
  year          = {2005},
  pages         = {271--310},
  volume        = {55},
  archiveprefix = {arXiv},
  doi           = {10.1146/annurev.nucl.55.090704.151526},
  eprint        = {nucl-ex/0502005},
}

@article{Baltz:2007kq,
    author = "Baltz, A. J.",
    editor = "Baur, G. and others",
    title = "{The Physics of Ultraperipheral Collisions at the LHC}",
    eprint = "0706.3356",
    archivePrefix = "arXiv",
    primaryClass = "nucl-ex",
    doi = "10.1016/j.physrep.2007.12.001",
    journal = "Phys. Rept.",
    volume = "458",
    pages = "1--171",
    year = "2008"
}

@Article{CMS:2020skx,
  author        = {Sirunyan, Albert M and others},
  journal       = {Phys. Rev. Lett.},
  title         = {{Observation of Forward Neutron Multiplicity Dependence of Dimuon Acoplanarity in Ultraperipheral Pb-Pb Collisions at $\sqrt{s_{NN}}$=5.02\,\,TeV}},
  year          = {2021},
  number        = {12},
  pages         = {122001},
  volume        = {127},
  archiveprefix = {arXiv},
  collaboration = {CMS},
  doi           = {10.1103/PhysRevLett.127.122001},
  eprint        = {2011.05239},
  primaryclass  = {hep-ex},
  reportnumber  = {CMS-HIN-19-014, CERN-EP-2020-196},
}

@Article{ALICE:2013wjo,
  author        = {Abbas, E. and others},
  journal       = {Eur. Phys. J. C},
  title         = {{Charmonium and $e^+e^-$ pair photoproduction at mid-rapidity in ultra-peripheral Pb-Pb collisions at $\sqrt{s_{\rm NN}}$=2.76 TeV}},
  year          = {2013},
  number        = {11},
  pages         = {2617},
  volume        = {73},
  archiveprefix = {arXiv},
  collaboration = {ALICE},
  doi           = {10.1140/epjc/s10052-013-2617-1},
  eprint        = {1305.1467},
  primaryclass  = {nucl-ex},
  reportnumber  = {CERN-PH-EP-2013-066},
}

@article{Klusek-Gawenda:2016euz,
    author = "K\l{}usek-Gawenda, Mariola and Lebiedowicz, Piotr and Szczurek, Antoni",
    title = "{Light-by-light scattering in ultraperipheral Pb-Pb collisions at energies available at the CERN Large Hadron Collider}",
    eprint = "1601.07001",
    archivePrefix = "arXiv",
    primaryClass = "nucl-th",
    doi = "10.1103/PhysRevC.93.044907",
    journal = "Phys. Rev. C",
    volume = "93",
    number = "4",
    pages = "044907",
    year = "2016"
}

@Article{Shao:2022stc,
  author        = {Shao, Ding Yu and Zhang, Cheng and Zhou, Jian and Zhou, Ya-Jin},
  journal       = {Phys. Rev. D},
  title         = {{Azimuthal asymmetries of muon pair production in ultraperipheral heavy ion collisions}},
  year          = {2023},
  number        = {3},
  pages         = {036020},
  volume        = {107},
  archiveprefix = {arXiv},
  doi           = {10.1103/PhysRevD.107.036020},
  eprint        = {2212.05775},
  primaryclass  = {hep-ph},
}

@Article{Hencken:2004td,
  author        = {Hencken, Kai and Baur, Gerhard and Trautmann, Dirk},
  journal       = {Phys. Rev. C},
  title         = {{Production of QED pairs at small impact parameter in relativistic heavy ion collisions}},
  year          = {2004},
  pages         = {054902},
  volume        = {69},
  archiveprefix = {arXiv},
  doi           = {10.1103/PhysRevC.69.054902},
  eprint        = {nucl-th/0402061},
}

@Article{Lin:2022flv,
  author        = {Lin, Shuo and Wang, Ren-Jie and Wang, JIan-Fei and Xu, Hao-Jie and Pu, Shi and Wang, Qun},
  journal       = {Phys. Rev. D},
  title         = {{Photoproduction of e+e- in peripheral isobar collisions}},
  year          = {2023},
  number        = {5},
  pages         = {054004},
  volume        = {107},
  archiveprefix = {arXiv},
  doi           = {10.1103/PhysRevD.107.054004},
  eprint        = {2210.05106},
  primaryclass  = {hep-ph},
}

@article{Jiang:2024vuq,
    author = "Jiang, Jun and Li, Shi-Yuan and Liang, Xiao and Liu, Yan-Rui and Qiao, Cong-Feng and Si, Zong-Guo and Yang, Hao",
    title = "{Pseudoscalar heavy quarkonium production in heavy ion ultraperipheral collision}",
    eprint = "2406.19735",
    archivePrefix = "arXiv",
    primaryClass = "hep-ph",
    month = "6",
    year = "2024"
}

\end{document}